\documentclass[%
aps,
prapplied,
superscriptaddress,
longbibliography,
 amsmath,amssymb,
 reprint,%
]{revtex4-2}

\usepackage{hyperref}

\usepackage[capitalise]{cleveref}

\usepackage{graphicx}
\usepackage{dcolumn}
\usepackage{bm}
\usepackage{float}
\usepackage{array}
\usepackage{textcomp}
\usepackage{physics}
\usepackage[dvipsnames]{xcolor}
\usepackage{multirow}
\usepackage{booktabs}
\usepackage{upgreek}
\usepackage{xr}
\usepackage{cleveref}
\usepackage{tabularx}
\usepackage{amsmath}
\usepackage{appendix}
\usepackage{gensymb}
\usepackage{fixltx2e}

\usepackage{newtxtext}
\usepackage{newtxmath}
\usepackage{longtable}

\usepackage[mathlines]{lineno}

\usepackage{multibib}
\newcites{supp}{Supplementary References} 

\newcolumntype{C}[1]{>{\centering\arraybackslash}p{#1}}

\begin{document}

\preprint{AIP/123-QED}

\title[]{Scaling Alternating-Bias-Assisted Annealing for Precision Transmon Frequency Targeting on Superconducting Quantum Processors}

\author{Xiqiao Wang}
\email{xwang@rigetti.com}
\affiliation{ 
Rigetti Computing, 775 Heinz Avenue, Berkeley, CA 94710, USA
}

\author{Mark Field}
\affiliation{ 
Rigetti Computing, 775 Heinz Avenue, Berkeley, CA 94710, USA
}

\author{Teng Zhang}
\affiliation{ 
Rigetti Computing, 775 Heinz Avenue, Berkeley, CA 94710, USA
}

\author{Xian Wu}
\affiliation{ 
Rigetti Computing, 775 Heinz Avenue, Berkeley, CA 94710, USA
}

\author{Ferhat Aydinoglu}
\affiliation{ 
Rigetti Computing, 775 Heinz Avenue, Berkeley, CA 94710, USA
}

\author{Joel Howard}
\affiliation{ 
Rigetti Computing, 775 Heinz Avenue, Berkeley, CA 94710, USA
}

\author{Angela Q. Chen}
\affiliation{ 
Rigetti Computing, 775 Heinz Avenue, Berkeley, CA 94710, USA
}

\author{Sara Elzeiny}
\affiliation{ 
Rigetti Computing, 775 Heinz Avenue, Berkeley, CA 94710, USA
}

\author{Robert Smith}
\affiliation{ 
Rigetti Computing, 775 Heinz Avenue, Berkeley, CA 94710, USA
}

\author{Timothy McSorley}
\affiliation{ 
Rigetti Computing, 775 Heinz Avenue, Berkeley, CA 94710, USA
}

\author{Nicholas Sharac}
\affiliation{ 
Rigetti Computing, 775 Heinz Avenue, Berkeley, CA 94710, USA
}

\author{Eyob Sete}
\affiliation{ 
Rigetti Computing, 775 Heinz Avenue, Berkeley, CA 94710, USA
}

\author{Alysson Gold}
\affiliation{ 
Rigetti Computing, 775 Heinz Avenue, Berkeley, CA 94710, USA
}

\author{Hilal Cansizoglu}
\affiliation{ 
Rigetti Computing, 775 Heinz Avenue, Berkeley, CA 94710, USA
}

\author{Greg Stiehl}
\affiliation{ 
Rigetti Computing, 775 Heinz Avenue, Berkeley, CA 94710, USA
}

\author{Josh Mutus}
\affiliation{ 
Rigetti Computing, 775 Heinz Avenue, Berkeley, CA 94710, USA
}

\author{Kameshwar Yadavalli}
\affiliation{ 
Rigetti Computing, 775 Heinz Avenue, Berkeley, CA 94710, USA
}

\author{Andrew Bestwick}
\affiliation{ 
Rigetti Computing, 775 Heinz Avenue, Berkeley, CA 94710, USA
}

\author{Stefano Poletto}
\affiliation{ 
Rigetti Computing, 775 Heinz Avenue, Berkeley, CA 94710, USA
}

\author{Raja Katta}
\affiliation{ 
Rigetti Computing, 775 Heinz Avenue, Berkeley, CA 94710, USA
}

\author{David P. Pappas}
\affiliation{ 
Rigetti Computing, 775 Heinz Avenue, Berkeley, CA 94710, USA
}

\date{August 25, 2026}

\begin{abstract}

Recent advances in the alternating-bias-assisted annealing (ABAA) technique have successfully mitigated intrinsic Josephson-junction (JJ) fabrication variations. This new technique enables precision qubit frequency tuning alongside simplicity. However, it is critical to enhance tuning throughput and yield while investigating the factors that drive targeting performance as the technology scales. Here, we characterize ABAA tuning performance within a 150-mm wafer process flow and extend this technique to simultaneous, multi-channel tuning, demonstrating that a wafer-scale JJ resistance tuning precision of $\sigma=0.50\pm0.05\%$ alongside a component-level yield of $\ge 98.8\%$ can be achieved. Furthermore, we demonstrate a strong correlation between yield, tuning speed, and junction breakdown voltage, establishing the latter as a vital process control parameter for meeting production goals. Finally, we demonstrate a successful implementation of ABAA tuning on a quad-module quantum processor (Rigetti Cepheus-1-36Q\texttrademark), where we achieve an empirical frequency targeting precision of $\sigma \sim 30\text{ MHz}$ in both qubit and qubit-qubit detuning frequencies, contributing to high median two-qubit gate fidelities. These results confirm the efficacy and scalability of ABAA for high-precision Hamiltonian targeting, a critical enabler for modular superconducting quantum processor technology.
\end{abstract}


\maketitle

\section{Introduction}

 Superconducting transmon qubits are a leading candidate for achieving utility-scale, fault-tolerant quantum computing \cite{PhysRevA.80.052312, saadatmand2024superconducting, mohseni2024build}, primarily due to their fast gate speeds and the viability of leveraging existing foundry fabrication and packaging techniques for scalability. Achieving high-fidelity quantum gate operations across large-scale superconducting quantum processors requires precise targeting of qubit frequencies. This precision is essential to realizing Hamiltonian designs that optimize device performance. First, it enables the targeting of desirable transmon detunings for two-qubit (2Q) gates -- such as minimizing leakage in adiabatic-CZ \cite{chen2026unlocking} or preserving dephasing times via shallow flux modulation in resonant-CZ \cite{Sete_para_2021}. Second, it mitigates crosstalk and leakage during one-qubit (1Q) operations by addressing frequency crowding in dense spectral allocations on large-scale processors. Furthermore, a precisely targeted qubit lattice significantly simplifies the automation of gate and measurement calibration on processors hosting hundreds of thousands of qubits.

Although scalable transmon fabrication techniques are available, the as-fabricated transmon targeting error is limited to approximately $3\%$ in resistance. This limitation stems from the extreme sensitivity of the Josephson junction resistance (and thus transmon inductance) to variations in junction area and effective barrier thickness on the nanometer and sub-nanometer scales, respectively \cite{Kreikebaum_2020,osman2021simplified, pishchimova2023improving}. Recent advances in post-fabrication Josephson-junction (JJ) tuning techniques \cite{granata2007localized, Hertzberg2021, LasiQScience2022, LasiQBerkeley2022, koppinen2007complete, migacz2003thermal, korshakov2024aluminum, balaji2024electronbeam, smirnov2025subangstrom}, which utilize various junction annealing interactions, have demonstrated significant promise in reducing these targeting errors. State-of-the-art tuning precision has been achieved by the laser-annealing technique, initially developed at IBM \cite{Hertzberg2021, LasiQScience2022}, and the alternating-bias-assisted annealing (ABAA) technique, pioneered at Rigetti \cite{pappas2024alternating, wang2024precision}. Both methods have demonstrated a frequency-equivalent tuning precision below $10\text{ MHz}$. Among these, the ABAA technique offers the additional distinct advantage of requiring only commercially available electrical test equipment for its implementation.

Since the discovery and first demonstration of transmon frequency targeting via the ABAA effect, the field has rapidly advanced in exploring annealing voltage sequence parameters \cite{Krizan_2026}, modeling the ABAA effect \cite{xkqs-grnd}, and elucidating the microscopic mechanisms of defect healing and performance enhancement in Josephson-junction-based qubits \cite{iaia2025non,WeigelHRTEMJofPhysCondMat2008}. However, foundry-scale implementation of ABAA on high-performance, scalable Quantum Processing Units (QPUs) has not been demonstrated.

In this paper, we demonstrate the scalability of ABAA throughput without compromising tuning precision through the implementation of simultaneous, multi-channel tuning. Furthermore, we characterize wafer-scale ABAA targeting and evaluate its temporal stability over a 14-week post-operation period. We explore the correlation between the ABAA short ratio, tuning speed time constants, and Josephson junction (JJ) breakdown voltage (BDV) over a range of annealing voltages, identifying the JJ BDV as a key fabrication process control parameter crucial for managing ABAA throughput and yield. Finally, we demonstrate a successful implementation of ABAA tuning on a quad-module quantum processor (Rigetti Cepheus-1-36Q\texttrademark) with high median 2-Q gate fidelity, achieving an empirical targeting precision of $\sim 30\text{ MHz}$ ($1\text{-}\sigma$) in both qubit frequency and detunings. We empirically observe a higher ceiling in 2Q gate performance as the qubit-qubit (q-q) detuning approaches the preferred regime specified by the Hamiltonian design, illustrating the clear gate performance benefits of tighter detuning targeting. Ultimately, our results validate the scalability of ABAA targeting on modular QPU architectures for realizing large, defect-free superconducting qubit lattices with high precision frequency allocation.

\begin{figure*}
\includegraphics[width=1.0\textwidth]{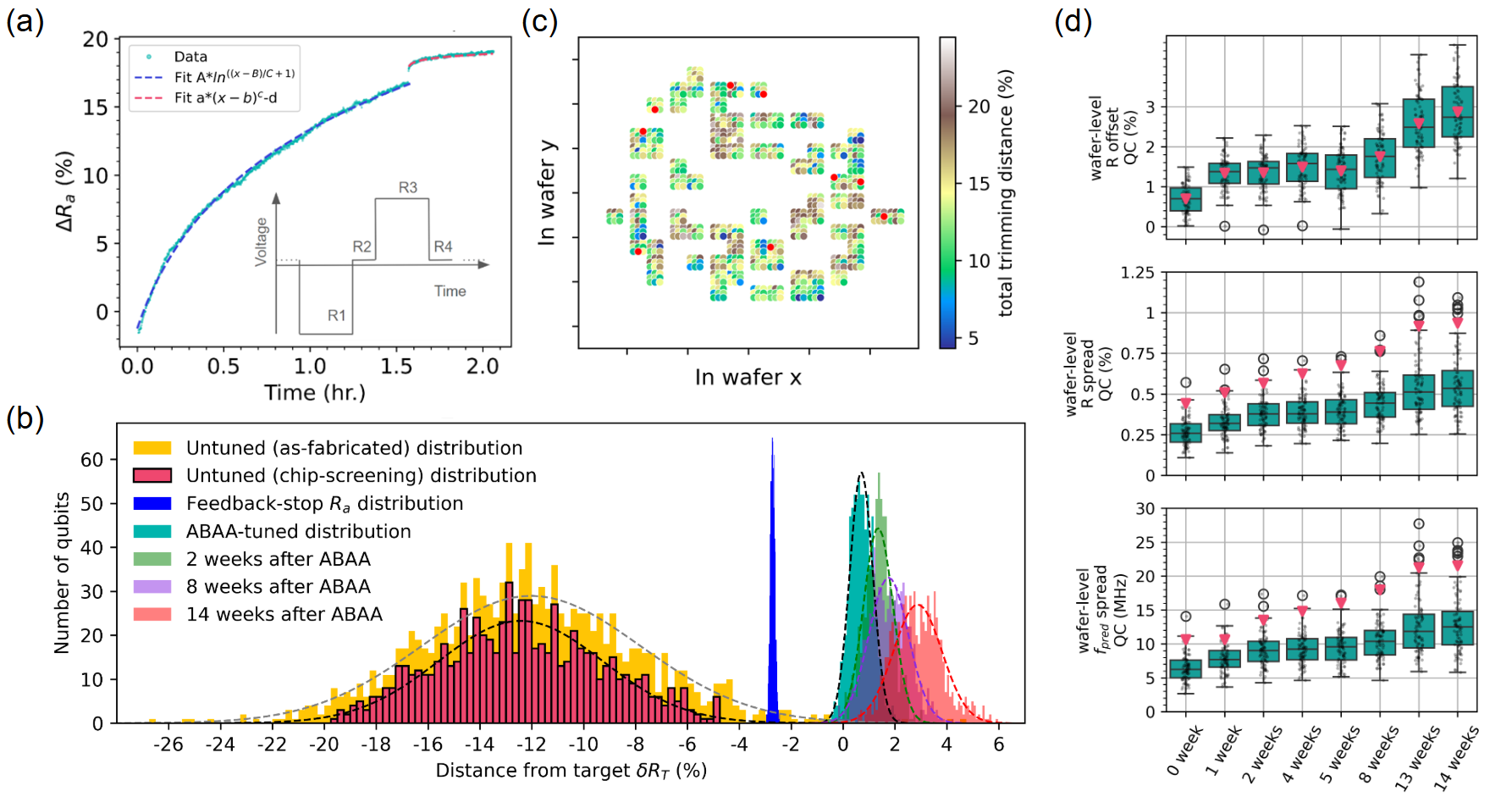}
\caption{ABAA targeting at the wafer scale. (a) A typical profile of junction resistance during the ABAA process. The blue and red dashed curves represent numerical fits to the active-tuning phase (where the alternating-bias annealing sequence is applied) and the post-tuning relaxation phase (where spontaneous relaxation in junction resistance is monitored at low bias), respectively. The inset illustrates the alternating-bias annealing voltage sequence used in this study, where steps $R_1$ and $R_3$ denote annealing intervals, and steps $R_2$ and $R_4$ denote monitoring intervals. (b) Histogram distributions of the Josephson junction (JJ) resistance, normalized to the target resistance, across different stages of the ABAA tuning workflow on a $150\text{-mm}$ wafer. As-Fabricated (Yellow): Initial resistance distribution for all JJs from the 136 nine-qubit (9Q) chiplets of a target design. Post-Screening (Pink): Distribution for JJs from the 83 9Q chiplets that passed the initial chip-screening process. Active-Tuning End (Blue): Monitor resistance at the conclusion of the active-tuning phase for each JJ. Final Tuned (Teal): ABAA-tuned JJ resistance measured immediately after tuning all chiplets that passed screening. This distribution excludes the 9 JJs that shorted out of the 747 total JJs tuned. The semi-transparent distributions to the right of the teal distribution illustrate the temporal aging at specific storage intervals, detailed further in (d). (c) A wafer map visualizing the ABAA-tuned components, where the color coding indicates the total tuning distance applied to each component, and components that shorted during the ABAA process are highlighted in red. (d) Temporal stability of the ABAA-targeted distribution monitored for over three months while the wafer was stored in a room-temperature, nitrogen-purged environment. Panels from top to bottom illustrate the time evolution of post-ABAA metrics normalized to the target: resistance offset, resistance spread, and the equivalent predicted frequency spread. Black data points represent statistics for individual 9Q chiplets, box plots summarize the chiplet-level statistics, and red triangles represent the aggregate statistics for the entire wafer.
}
\label{Fig:Figure1}
\end{figure*}

\section{ABAA at the wafer scale}

We implemented the room-temperature ABAA targeting protocol as previously described \cite{pappas2024alternating, wang2024precision} at the wafer scale. Each ABAA annealing voltage cycle consists of four steps (inset of Figure 1(a)): the first and third steps apply opposite-polarity annealing voltages ($V_a$), while the second and fourth steps apply a monitoring voltage ($V_m \ll V_a$) to track the junction resistance within the small-bias linear $I\text{-}V$ regime, providing real-time feedback of the junction resistance after each annealing pulse. 

Figure 1(a) shows a typical tuning profile comprising two distinct phases: (1) active tuning, where the annealing sequence drives a logarithmic increase in resistance over time (or with the number of annealing cycles), and (2) post-tuning relaxation, where the resistance continues to follow a power-law increase after the final annealing pulse. Targeted ABAA requires calibrating this post-tuning relaxation to precisely terminate the annealing sequence once the resistance monitored during active tuning ($R_a$) reaches a predetermined threshold, thereby allowing the junction to relax to the final target resistance ($R_T$) over a defined relaxation or aging window \cite{wang2024precision}.

The implementation of wafer-scale ABAA targeting is illustrated using an example involving 136 9-qubit (9Q) chiplets of a target design on a 150-mm wafer (Fig. 1(b)). The as-fabricated Josephson junction (JJ) resistance spread across the wafer exhibits a $1\text{-}\sigma$ variation of $\pm 4\%$ (yellow distribution in Fig. 1(b)), which is limited by fabrication precision. An average global offset of approximately $-12\%$ relative to the target is intentionally introduced to accommodate ABAA tuning. 

Prior to ABAA, a chip-screening step is performed to select chiplets suitable for tuning. In general, chiplets are filtered out if they contain components that are either: (1) too close to the target resistance $R_T$ from below ($< 3\%$), leaving insufficient headroom for post-tuning relaxation; (2) above the target resistance $R_T$, as we only tune the resistance in the upward direction; or (3) excessively far below the target resistance $R_T$ (e.g., $> 25\%$ below), in light of the logarithmic decrease in tuning speed over distance (Fig. 1(a), blue dashed line), would necessitate an impractically long tuning duration. In this example, 83 chiplets passed screening, forming a distribution of $\mu\pm\sigma=-12 \pm 3\%$ (red in Fig. 1(b)) that proceeded to the ABAA process. 

The narrow blue histogram in Fig. 1(b) represents the live-feedback monitor resistance at the termination of the annealing voltage sequence for each qubit. This distribution is centered at $-2.7\%$ (determined by targeting calibration) with a spread of $\sigma\ll 0.1\%$, bounded fundamentally by the incremental resistance step size during ABAA. After all 83 chiplets complete the ABAA process and are reprobed, the resulting resistance distribution yields a typical $1\text{-}\sigma$ spread of $\pm 0.4\%$ and a global offset of $0.7\%$ from the target (teal histogram) in this example. 

Two main wafer-scale effects, which may become more pronounced on larger wafer sizes, contribute to this targeting spread and to wafer-to-wafer offset errors; (1) Intrinsic component variation: Even under identical tuning and relaxation conditions, individual components exhibit variations in relaxation rates. This is attributed to spatial variations in microstructure-related junction/barrier dynamics across the wafer.(2) Variations in relaxation time: The total duration of the ABAA tuning run dictates that the time available for post-tuning relaxation prior to final reprobing can range from hours to days between the first and last completed components on a wafer, depending on the total required tuning allocation. Increasing the number of parallel ABAA channels, as will be discussed in Section III, provides an effective strategy to mitigate this wafer-level relaxation time-span variation.

Fig. 1(c) presents a wafer map of the components tuned in Fig. 1(b), color-coded by the total tuning distance as a percentage. In total, approximately 1.6 million alternating annealing voltage steps were consumed to tune the 747 qubit components to the target resistance (comprising the 83 chiplets that passed screening, with 9 qubits per chiplet). In this example, 9 qubits across 9 distinct chiplets shorted before reaching their tuning target during ABAA. This corresponds to a yield of 89\% at the chiplet level, or 98.8\% at the component level. The probability of short per annealing pulse is $\sim6 \times 10^{-6}$. Here, we note that short rates at the chiplet and component levels serve as useful metrics for quantifying product-delivery yield, whereas the pulse-level short ratio is relatively independent of tuning-distance and targeting factors, making it more suitable for evaluating the intrinsic yield of the ABAA process. As illustrated in more detail in Section IV, the short ratio can be further suppressed by reducing the amplitude of the annealing voltage, while the associated increase in tuning time across components can be offset by increasing the number of simultaneous ABAA channels.

Although the $98.8\%$ component-level yield in this particular example is insufficient to realize a defect-free monolithic QPU beyond the 100-qubit scale, the modular architecture employed in this study circumvents this limitation. Specifically, a much larger, defect-free qubit lattice can be constructed by tiling smaller, short-free chiplets. Assuming a maximum capacity of 212 9Q chiplets on a single 150-mm wafer, and applying the ABAA-limited chiplet-level yield of $89\%$ from this example, we project a maximum yield of 188 short-free 9Q chiplets. These can be interconnected into a defect-free lattice comprising $> 1600$ qubits.

The stability of the ABAA-tuned resistance spread was monitored for over three months while storing the wafer in a nitrogen-purged ($\text{N}_2$), room-temperature environment. The semi-transparent distributions in Fig. 1(b) illustrate the temporal evolution of the initial post-ABAA distribution (teal). By the week-14 post-operation, the wafer-level distribution exhibited a global aging offset of approximately $2\%$, and the $1\text{-}\sigma$ spread widened from $0.4\%$ to approximately $0.9\%$. 

Figure 1(d) provides a more quantitative comparison of these wafer-level post-ABAA aging indicators (red triangles) against chip-level tuning statistics (black dots and boxes). While the wafer-level aging offset aligns consistently with the chip-level median offset, the total wafer-level spread is approximately double the chip-level median spread. This discrepancy is primarily driven by the spatial variation in chip-to-chip offsets across the wafer substrate. Initially (week-0), the chip-level median resistance spread was $0.25\%$ (corresponding to $\sim 6\text{ MHz}$ in predicted frequency), compared to a total wafer-level spread of $0.5\%$ ($\sim 11\text{ MHz}$). By week-14, these values increased to $\sim 13\text{ MHz}$ and $\sim 23\text{ MHz}$, respectively. (Details on the frequency prediction method are provided in Supplementary Material Appendix A).

This temporal broadening of the resistance spread constitutes a major source of error in the frequency-targeting workflow (see Section V). Consequently, maximizing the benefits of ABAA requires minimizing wait-time variations between tuning and cryogenic deployment, ideally through on-demand ABAA. Alternatively, cryogenic storage offers a potential solution for extending the shelf life of ABAA-tuned components \cite{budoyo2026characterization, Krizan_2026}.

\begin{center}
\begin{longtable}{|l|l|l|}
\caption{JJ resistance distributions in Figure 1} \label{tab:long} \\

\hline \multicolumn{1}{|c|}{\textbf{Steps}} & \multicolumn{1}{c|}{\textbf{$\langle R\rangle$ from $R_T$}} & \multicolumn{1}{c|}{\textbf{$\sigma$}} \\ \hline 
\endfirsthead

\multicolumn{3}{c}%
{{\bfseries \tablename\ \thetable{} -- continued from previous page}} \\
\hline \multicolumn{1}{|c|}{\textbf{First column}} & \multicolumn{1}{c|}{\textbf{Second column}} & \multicolumn{1}{c|}{\textbf{Third column}} \\ \hline 
\endhead

\hline \multicolumn{3}{|r|}{{Continued on next page}} \\ \hline
\endfoot

\hline \hline
\endlastfoot

As-fabricated & -12.0\% & 4.2\% \\
Chip-screening & -12.4\% & 3.2\% \\
Feedback Tuning $R_a$ & -2.7\% & \textless0.1\% \\
ABAA completed & 0.70\% & 0.44\% \\
Week-1 after ABAA & 1.34\% & 0.51\% \\
Week-2 after ABAA & 1.35\% & 0.57\% \\
Week-4 after ABAA & 1.49\% & 0.62\% \\
Week-5 after ABAA & 1.40\% & 0.68\% \\
Week-8 after ABAA & 1.75\% & 0.76\% \\
Week-13 after ABAA & 2.58\% & 0.92\% \\
Week-14 after ABAA & 2.87\% & 0.94\% \\
\end{longtable}
\end{center}

\begin{figure*}
\includegraphics[width=1.0\textwidth]{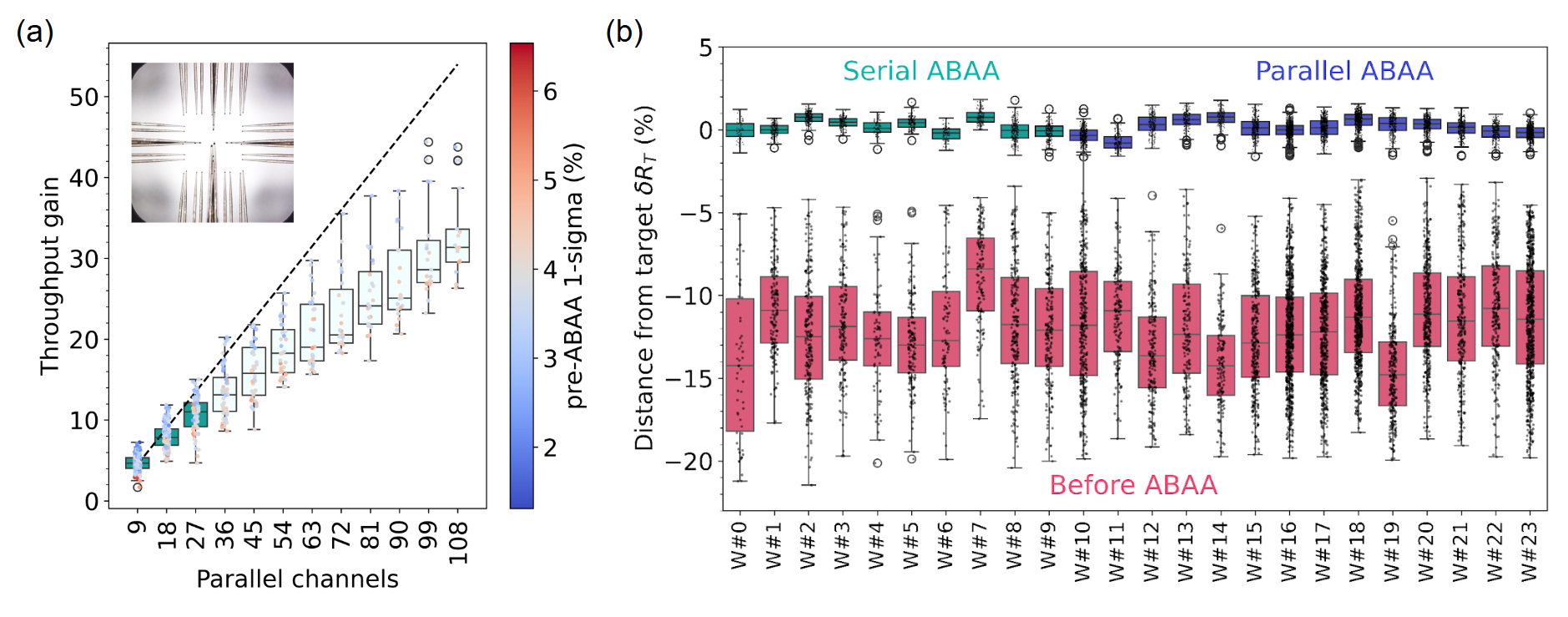}
\caption{Parallel ABAA scaling and spread reduction performance. (a) Throughput gain plotted as a function of the number of parallel channels. Each data point represents the throughput gain per probe-card touchdown, with the marker color indicating the initial pre-tuning spread. Experimental data were collected for $9$, $18$, and $27$ parallel channels (teal boxplots), while projections for higher channel counts were estimated via extrapolation from the existing datasets (white boxplots). Inset: An optical image of a cantilever-probe array used in this study. (b) Comparison of wafer-scale resistance spread reduction between the serial ABAA (teal) and parallel ABAA (neon blue) methods. For each wafer index, the lower and upper boxes display the resistance distributions relative to the tuning target, evaluated pre- and post-ABAA tuning, respectively. The wafer-level resistance targeting statistics derived from the data in Fig. 2(b) are summarized in Table II. }
\label{Fig:Figure2}
\end{figure*}

\section{Parallel ABAA} 

The throughput of the ABAA process, under a given device condition and ABAA recipe, can be significantly increased through simultaneously tuning multiple components in parallel (parallel ABAA). This is accomplished by utilizing multi-channel source measure units (SMUs) and high-density probe cards, which are technologies already mature within the semiconductor industry \cite{weeden2003probe}. The inset in Fig. 2(a) shows an optical image of a representative 27-channel cantilever-type probe card used in this study. This system enables simultaneous ABAA tuning of 27 components (qubits plus tunable couplers (TCs) and inter-modular couplers (IMCs)) within a single probe-card touchdown, where the total touchdown duration is dictated by the component requiring the longest tuning time. Consequently, the throughput gain from this parallel operation is quantified by dividing the aggregate tuning time of all 27 components by the duration of the single most time-consuming component.

Fig. 2(a) illustrates how this throughput gain scales with the number of parallel channels. Data points for 9, 18, and 27 parallel channels were collected experimentally, while behavior at higher channel counts was extrapolated from the existing datasets. At 9 parallel channels, the median throughput gain is $\sim 4.5$ (approximately half the parallel channel count). Within this distribution, as indicated by the data point color, the gain also exhibits a negative correlation with the pre-tuning spread, consistent with expectations from a statistical point of view. 

As the number of parallel channels increases, the throughput gain increasingly deviates below the $y = 0.5x$ guideline (dashed line in Fig. 3(a)). This sub-linear scaling occurs because the influence of the single most time-consuming component becomes more pronounced, forcing a larger fraction of components to remain idle during a single touchdown. Despite this nonlinearity, projections indicate that implementing 108 parallel channels for ABAA, which is equivalent to a 36Q QPU (36 qubits plus TCs and IMCs) in a single probe card touchdown, could still achieve an approximate 30-fold gain in throughput.

\begin{center}
\begin{longtable}{|l|l|l|}
\caption{Wafer-level resistance targeting performance in Figure 2} \label{tab:long} \\

\hline \multicolumn{1}{|c|}{\textbf{Wafer-level statistics}} & \multicolumn{1}{c|}{\textbf{Serial ABAA}} & \multicolumn{1}{c|}{\textbf{Parallel ABAA}} \\ \hline 
\endfirsthead

\multicolumn{3}{c}%
{{\bfseries \tablename\ \thetable{} -- continued from previous page}} \\
\hline \multicolumn{1}{|c|}{\textbf{Wafer-level statistics}} & \multicolumn{1}{c|}{\textbf{Serial ABAA}} & \multicolumn{1}{c|}{\textbf{Parallel ABAA}} \\ \hline 
\endhead

\hline \multicolumn{3}{|r|}{{Continued on next page}} \\ \hline
\endfoot

\hline \hline
\endlastfoot

Number of wafers & 10 & 14 \\
Pre-ABAA $\langle R\rangle$ from $R_T$ & $-11.4 \pm 1.4\%$ & $-12.1 \pm 1.2\%$ \\
Pre-ABAA $\sigma$ & $3.7 \pm 0.5\%$ & $3.3 \pm 0.4\%$ \\
Post-ABAA $\langle R\rangle$ from $R_T$ & $0.07 \pm 0.32\%$ & $0.13 \pm 0.38\%$ \\
Post-ABAA $\sigma$ & $0.53 \pm 0.06\%$ & $0.50 \pm 0.05\%$ \\
\end{longtable}
\end{center}

Fig. 2(b) compares the resistance spread reduction for wafers tuned via serial (teal) versus parallel (blue) ABAA. For each wafer, the lower and upper boxes represent the pre- and post-tuning resistance distributions, respectively, with each data point corresponding to a individual qubit. Table II summarizes the wafer-level statistics, demonstrating that parallel and serial ABAA achieve equivalent targeting performance. Furthermore, the significantly higher density of data points provided by parallel-ABAA wafers underscores the technique's ability to tune more components within standard production-line time constraints.

\begin{figure*}
\includegraphics[width=0.9\textwidth]{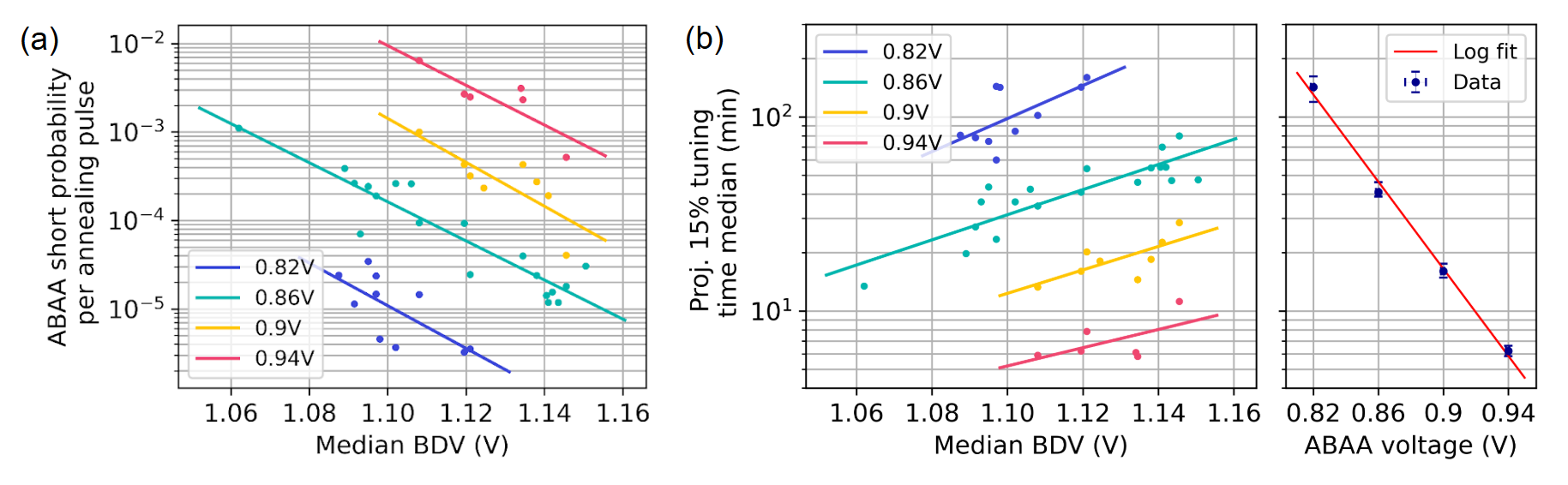}
\caption{ Relationship between the Josephson junction (JJ) breakdown voltage and the yield and throughput of the ABAA process. (a) Wafer-level ABAA short probability per annealing pulse versus wafer-median breakdown voltage (BDV) across a selection of alternating annealing amplitudes. (b) Left panel: Wafer-median ABAA tuning time (at a $15\%$ active tuning) versus wafer-median BDV across the same selection of annealing amplitudes. Right panel: Wafer-median ABAA tuning time (at a $15\%$ active tuning) versus ABAA annealing voltage on a wafer with a median breakdown voltage (BDV) of 1.12 V. Error bars represent the 25\textsuperscript{th} to 75\textsuperscript{th} percentiles. In both (a) and (b), each data point corresponds to the statistical results from a single $150\text{-mm}$ wafer. The wafer-median BDV is calculated as the average of the median positive and median negative BDVs measured on a dedicated set of BDV test qubit devices prior to the ABAA process. The ABAA short probability per annealing pulse is defined as the total number of short occurrences observed during the tuning process divided by the total number of applied annealing pulses. The ABAA tuning time is determined by extrapolating the required duration to a $15\%$ tuning distance based on a logarithmic fit of the active tuning profile, as illustrated in Fig. 1(a).}
\label{Fig:Figure3}
\end{figure*}

\section{ABAA yield and throughput control via breakdown voltage monitoring}

In this study, we introduce two metrics that are independent of the total tuning distance to quantify the tuning speed and short probability: \textit{Tuning speed} is quantified by fitting the active tuning profile with a logarithmic function (as shown in Fig. 1(a)) and extrapolating the tuning time required to achieve a preset $15\%$ change in resistance. The \textit{short probability} is defined as the number of breakdown occurrences per alternating annealing voltage pulse.

The tuning speed and short probability during ABAA depend heavily on both the specific annealing voltage sequence parameters and the intrinsic properties of the junction and barrier. In general, a faster tuning speed increases the likelihood of junction failure (i.e., a dielectric breakdown leading to an electrical short) during the tuning process. Consequently, optimizing ABAA throughput requires carefully balancing the competing goals of tuning speed and process yield. 

Figure 3 summarizes the room-temperature experimental data collected across a set of commonly used annealing voltage amplitudes, characterizing the dependence of both the wafer-scale ABAA short probability and the median tuning time (at a fixed $15\%$ active tuning distance) on the wafer-median breakdown voltages. As illustrated in Fig. 3(a), the ABAA short probability per annealing pulse decreases exponentially with increasing BDV at a constant annealing voltage, as well as when the annealing voltage amplitude is reduced (moving from red to blue). The extreme sensitivity of the short probability to the baseline BDV is noteworthy; for example, along the $0.86\text{ V}$ operational line, a minor $\sim 100\text{ mV}$ ($\sim 9\%$) decrease in BDV (from $1160\text{ mV}$ to $1060\text{ mV}$) leads to an increase in the short probability by over two orders of magnitude. Turning to Fig. 3(b) left-panel, the $15\%$ tuning time (wafer-median) exhibits exponential growth as a function of increasing BDV at a constant annealing voltage. Conversely, reducing the annealing voltage amplitude also results in an exponential increase in the required tuning duration, as illustrated in the right panel of Fig. 3(b) for a wafer with a median BDV of $\sim1.12V$.  

The positive correlation between the JJ BDV and the $\mathit{RA}$ product is well-established \cite{zeng2015direct, FritzStructureMechancialAlOxPhysRevMater2019}, with both serving as crucial metrics for evaluating the quality of the JJ barrier, electrodes, and interface. These properties can be precisely targeted by adjusting various fabrication parameters, most directly through optimization of the barrier oxidation process \cite{zeng2015direct}. 

Achieving predictable ABAA yield and throughput necessitates effective control and monitoring of critical Josephson junction (JJ) properties. The phase diagrams presented in Fig. 3 highlight the JJ BDV as a key process control knob and a predictive indicator of ABAA speed and yield at given tuning conditions. We implement an approach wherein the wafer-median BDV is first evaluated on a pre-defined set of test chips to guide the selection of the annealing voltage for tuning actual components on the same wafer. This methodology aids in meeting production delivery goals while providing resilience against variations in upstream fabrication processes and materials. In the event of a drift in baseline JJ properties or BDV, these phase diagrams allow for small adjustments to the annealing voltage, counteracting device variations to maintain ABAA yield and throughput within the targeted operating window (see Supplementary Material Appendix B). Finally, it is worth noting that careful consideration must be given to the corresponding adjustment of the relaxation calibration, as it depends on both the applied annealing voltage and the baseline barrier properties\cite{wang2024precision}.

\begin{figure*}
\includegraphics[width=\textwidth]{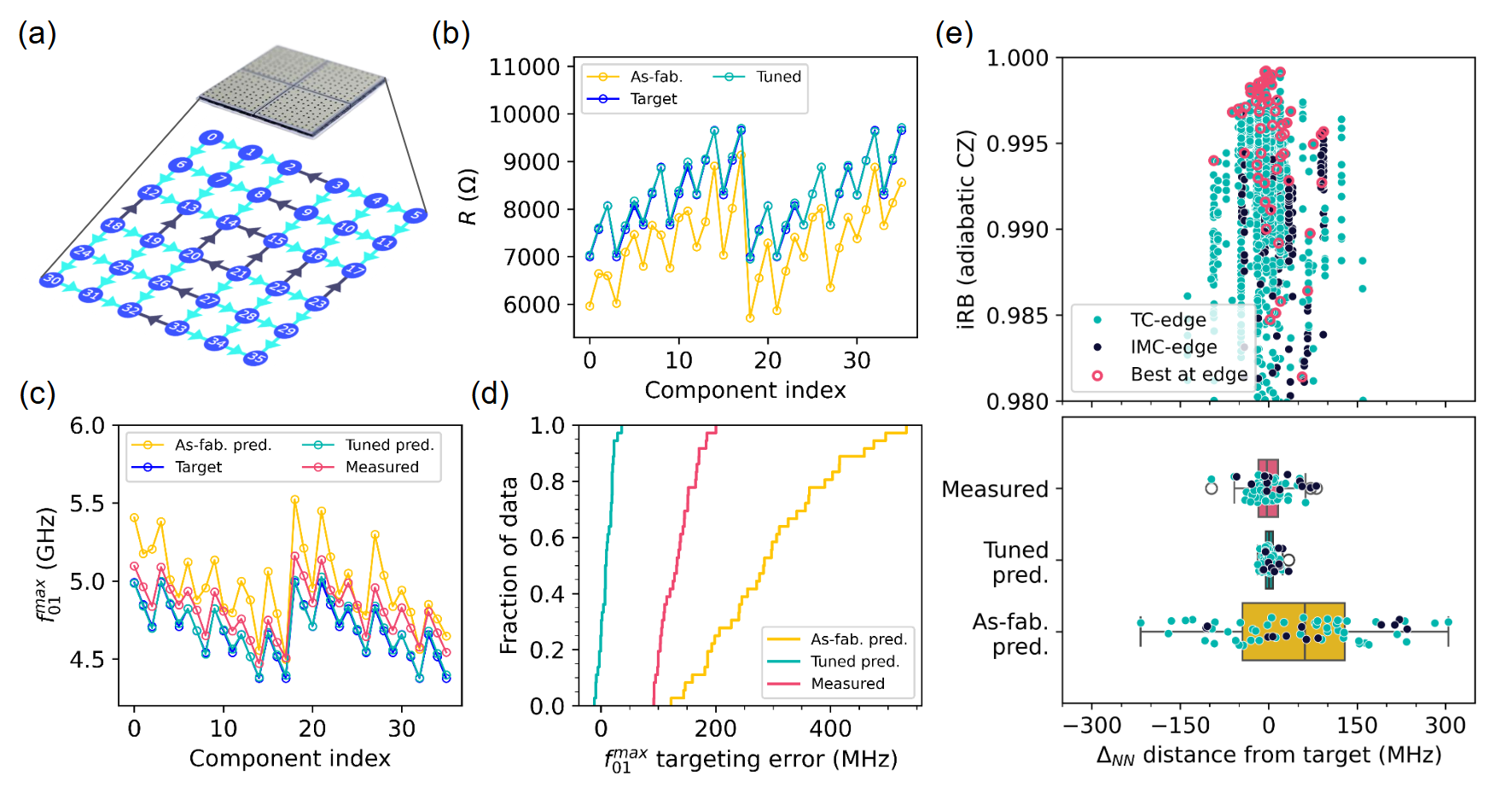}
\caption{Performance of resistance, frequency, and detuning targeting on a Rigetti Cepheus-1-36Q\texttrademark modular quantum processor. (a) Top: A photograph of a packaged Cepheus-1-36Q\texttrademark quantum processor. Bottom: A qubit lattice diagram of the processor, composed of four interconnected 9Q chiplets. Blue circles represent individual qubits, labeled with their corresponding component indices. Qubits within each chiplet are connected by tunable coupler (TC) edges (cyan), while those on adjacent chiplets are linked by inter-module coupler (IMC) edges (black). (b) Comparison of the target resistance (blue) with the probed resistance before (yellow) and after (teal) ABAA tuning. (c) Comparison of the Hamiltonian design (blue) and experimentally measured (pink) qubit $f_{01}^{\rm max}$, alongside predicted $f_{01}^{\rm max}$ values derived from room-temperature probed resistances pre-ABAA (yellow) and post-ABAA (teal). (d) Empirical cumulative distribution function (ECDF) plot comparing $f_{01}^{\rm max}$ targeting errors between the experimentally measured distribution (pink) and the predictedn distributions before (yellow) and after (teal) ABAA tuning. (e) Lower panel: nearest-neighbor qubit-qubit $f_{01}^{\rm max}$ detuning ($\Delta_{NN}$) targeting error relative to the design. From bottom to top, the distributions represent frequencies predicted before ABAA tuning, predicted after ABAA tuning, and measured experimentally. Teal and black markers denote data points from tunable coupler (TC) edges and inter-module coupler (IMC) edges, respectively. Adiabatic-CZ 2Q gate fidelity, benchmarked via interleaved randomized benchmarking (iRB) across all 48 TC edges and 11 live IMC edges, plotted as a function of the q-q detuning ($\Delta_{NN}$) deviation from the design value during gate activation. Pink circles highlight the maximum achieved 2Q fidelity for each nearest-neighbor 2Q edge.
}
\label{Fig:Figure4}
\end{figure*}


\section{Frequency targeting performance on modular quantum processors}

Finally, we experimentally demonstrate the successful implementation of ABAA-assisted Hamiltonian targeting on the Rigetti Cepheus-1-36Q\texttrademark quantum processor. First deployed for public access in August 2025, the processor achieved a median two-qubit gate (CZ) fidelity of 99.5\%. 

As illustrated in Figure 4(a), the Cepheus-1-36Q\texttrademark processor features a modular architecture \cite{gold2021entanglement, field2024modular} comprising four interconnected 9Q chiplets. In the schematic, the square lattice of blue circles represents the 36-qubit array, with individual qubits labeled at each node. Nearest-neighbor qubits within each chiplet are connected by tunable coupler (TC) edges (cyan), while those on adjacent chiplets are linked by inter-module coupler (IMC) edges (black). Arrows on each edge point from the higher-frequency to the lower-frequency qubit. In the context of the tunable transmon qubits used in this study, the qubit frequency refers to the maximum $|0\rangle\rightarrow |1\rangle$ transition frequency at zero flux bias. Here, $f_{01}^{\rm max}$ denotes the dressed qubit frequency, characterized while the TC and IMC idle at $ZZ=0$. 

To achieve high adiabatic CZ two-qubit gate fidelities, our Hamiltonian design optimizes the nearest-neighbor qubit-qubit $f_{01}^{\rm max}$ detuning, denoted as $\Delta_{NN}$. Specifically, the detunings along the tunable coupler (TC) and inter-module coupler (IMC) edges are designed such that $\Delta_{NN}^{TC} \approx \alpha_q / 2$ and $\Delta_{NN}^{IMC} \approx 3\alpha_q / 2$, where $\alpha_q$ is the qubit anharmonicity. For both edge types, the effective adiabatic-CZ detuning—defined as $|\alpha_q - \Delta_{NN}|$—is approximately $\alpha_q / 2$. This regime is highly desirable as it minimizes both gate leakage and flux-noise-induced decoherence in tunable transmon qubits \cite{chen2026unlocking}. Additionally, the design incorporates staggered $\Delta_{NN}$ patterns to prevent frequency collisions between next-nearest-neighbor qubits. The resulting Hamiltonian design values for $f_{01}^{\rm max}$ and their corresponding resistance targets are illustrated by the blue line profiles in Fig. 4(c) and (b), respectively.

Fig. 4(b) overlays the target qubit resistance with the room-temperature probed resistances at two stages: as-fabricated and post-ABAA tuning. The as-fabricated resistance distribution exhibits a deviation of $\mu \pm \sigma = -11.32 \pm 3.55\%$ relative to the tuning target (summarized in Table II). This intentional global offset $\mu$ is designed to facilitate ABAA tuning, while the spread $\sigma$ is limited by JJ fabrication variations. Following the completion of wafer-level ABAA tasks, the tuned resistance distribution was immediately probed, yielding a deviation of $\mu \pm \sigma = 0.39 \pm 0.49\%$ from the target. In the week following this post-ABAA probing, the tuned chips underwent wafer dicing, flip-chip bonding, and package assembly before being loaded into a dilution refrigerator for cool-down characterization, with no further resistance measurements taken.

Figure 4(c) compares the design $f_{01}^{\rm max}$ against both the experimentally measured $f_{01}^{\rm max}$ and the values predicted from the as-fabricated and ABAA-tuned resistances. Figure 4(d) plots the corresponding cumulative distributions of the $f_{01}^{\rm max}$ targeting errors across these three stages. Consistent with the narrowing of the resistance distributions, ABAA tuning yields a $\sim 9$-fold reduction in the predicted frequency spread. Specifically, this example demonstrates an improvement from an as-fabricated error distribution of $\mu \pm \sigma = 294 \pm 106\text{ MHz}$ to an ABAA-tuned distribution of $\mu \pm \sigma = 9 \pm 12\text{ MHz}$. Following cooldown and experimental characterization at base temperatures ($\sim 10\text{ mK}$), the measured $f_{01}^{\rm max}$ targeting error was found to be $\mu \pm \sigma = 132 \pm 30\text{ MHz}$. Although this still represents a significant improvement over the as-fabricated distribution, a noticeable increase in both the global offset and spread is observed relative to the post-ABAA predictions. The exact origins of these discrepancies remain under investigation; however, they likely result from a combination of two factors: (1) global offset and resistance variations accosiated with pre-cooldown aging and thermal cycling, and (2) systematic inaccuracies in frequency prediction modeling and calibration. Consequently, the measured distribution serves as an empirical upper bound for the cumulative effect of these error sources.

Because targeting the correct qubit-qubit detuning ($\Delta_{NN}$) is critical for optimal two-qubit gate performance, it serves as a primary driving criterion for our Hamiltonian design. In the lower panel of Fig. 4(e), we plot the $\Delta_{NN}$ targeting errors on the Cepheus-1-36Q\texttrademark processor at the three relevant stages: as-fabricated, post-ABAA tuning, and experimental measurement. Following ABAA tuning, the predicted $\Delta_{NN}$ error reduces significantly, improving from an as-fabricated distribution of $\mu \pm \sigma = 51 \pm 123\text{ MHz}$ to an ABAA-tuned distribution of $\mu \pm \sigma = 2 \pm 11\text{ MHz}$. The experimentally measured $\Delta_{NN}$ targeting error is $\mu \pm \sigma = 1 \pm 32\text{ MHz}$. It is worth highlighting that, while the measured $\Delta_{NN}$ spread increases to $\sim 32\text{ MHz}$ (up from the predicted $11\text{ MHz}$ post-ABAA) due to the increased spread in $f_{01}^{\rm max}$, the global offset error in $\Delta_{NN}$ remains negligible. This demonstrates that the relative frequencies between qubits are preserved, as common global offset errors in absolute $f_{01}^{\rm max}$ cancel out when deriving $\Delta_{NN}$. Furthermore, because all 9Q chiplets share the same level of global offset in $f_{01}^{\rm max}$, we observe no significant difference in detuning targeting errors between the TC edges ($\Delta_{NN}^{TC}$) and the IMC edges ($\Delta_{NN}^{IMC}$).

In the upper panel of Fig. 4(e), we plot the adiabatic-CZ two-qubit gate fidelity—benchmarked via interleaved randomized benchmarking (iRB) across all 48 TC edges and 12 IMC edges—as a function of the qubit-qubit detuning deviation from the design target during gate activation. This plotted detuning range closely tracks the measured $\Delta_{NN}$ error distributions shown in the lower panel. Red circles highlight the maximum fidelity achieved along each edge. Consistent with the aforementioned design considerations and their influence on gate execution details, we observe a general trend: higher maximum gate fidelities are attainable for $\Delta_{NN}$ values that are ABAA-tuned closer to their design targets. This provides empirical evidence for the performance benefits of realizing optimized Hamiltonian designs with high precision and accuracy, further motivating our efforts to push the boundaries of frequency targeting on large-scale quantum processors.

\begin{center}
\begin{longtable*}{|l|l|l|l|}
\caption{Resistance and frequency targeting performance in Cepheus-1-36Q\texttrademark} \label{tab:long} \\

\hline \multicolumn{1}{|c|}{\textbf{Cepheus-1-36Q\texttrademark}} & \multicolumn{1}{c|}{\textbf{R targeting $\mu\pm\sigma$}} & \multicolumn{1}{c|}{\textbf{$f_{01}^{\rm max}$ targeting $\mu\pm\sigma$}} & \multicolumn{1}{c|}{\textbf{$\Delta_{NN}$ targeting $\mu\pm\sigma$}} \\ \hline 
\endfirsthead

\multicolumn{4}{c}%
{{\bfseries \tablename\ \thetable{} -- continued from previous page}} \\
\hline \multicolumn{1}{|c|}{\textbf{First column}} & \multicolumn{1}{c|}{\textbf{Second column}} & \multicolumn{1}{c|}{\textbf{Third column}} & \multicolumn{1}{c|}{\textbf{Fourth column}} \\ \hline 
\endhead

\hline \multicolumn{4}{|r|}{{Continued on next page}} \\ \hline
\endfoot

\hline \hline
\endlastfoot

Before ABAA & $-11.32\pm3.55\%$ & $294\pm106MHz$ & $51\pm123MHz$ \\
After ABAA & $0.39\pm0.49\%$ & $9\pm12MHz$ & $2\pm11MHz$ \\
$f_{01}^{\rm max}$ meas. & - & $132\pm30MHz$ & $1\pm32MHz$ \\
\end{longtable*}
\end{center}

\section{Conclusion}

In summary, we have demonstrated the scalability of ABAA technology for precision Hamiltonian targeting on superconducting quantum processors, paving the way for its foundry implementation at the wafer scale. By integrating industry-standard probe-card technology, we performed simultaneous ABAA on multiple qubits in parallel, achieving a wafer-level Josephson junction (JJ) resistance targeting precision of $\sigma = 0.50 \pm 0.05\%$. Notably, we project a 30-fold increase in ABAA throughput by expanding the number of parallel channels from 1 to 108. Longitudinal tracking of an ABAA-tuned wafer reveals reasonable temporal stability, with wafer-level spread degradation constrained to a factor of $\sim 2$ for up to 14 weeks post-tuning. Furthermore, by demonstrating the exponential sensitivity of key ABAA yield and throughput metrics (shorting probability and tuning speed) to the wafer-median breakdown voltage (BDV) of JJ components, we identified BDV as a critical fabrication process control parameter for meeting production goals. Crucially, we achieved a 98.8\% component-level short-free yield during tuning at the 150-mm wafer scale. This high yield could potentially allow for the tiling of a short-free lattice comprising over 1600 qubits using 9Q chiplets tuned from a single wafer.

Finally, we experimentally analyzed the sources of resistance and frequency targeting errors and demonstrated the successful implementation of ABAA targeting on Rigetti's modular Cepheus-1-36Q\texttrademark{} quantum processor. This implementation achieved an empirical targeting precision of $\sim 30\text{ MHz}$ ($1\sigma$) relative to an optimized Hamiltonian design in both $f_{01}^{\rm max}$ and qubit-qubit detunings, contributing directly to the high two-qubit gate fidelities achieved on the processor.

Future work to continually improve Hamiltonian targeting precision and accuracy includes: (1) refining frequency prediction models; (2) improving the understanding and calibration of JJ resistance variations during pre-cooldown processing; and (3) mitigating wafer-to-wafer global offset errors when interconnecting chiplets from multiple wafers into large-scale, modular quantum processors.

\section*{acknowledgments}
We thank the Rigetti Fab-1 team for junction process development and sample fabrication.

\section*{contributions}
X. Wang conceived and implemented the ABAA targeting at scale and performed the experiments; M. F. and X. Wang developed the parallel ABAA concept and hardware setups; J. H., T. Z., and A. G. developed the frequency prediction framework; N. S., F. A., and H. C. developed the JJ fabrication process; X. Wu, S. P., A. Q. C., and E. S. designed the Hamiltonian and performed qubit frequency and gate characterizations; S. E., R. S., X. Wang., and T. M. developed the software infrastructure; R. K., J. M., K. Y., A. B., G. S., and D. P. P. provided technical guidance and management leadership. X. Wang analyzed the data and wrote the manuscript with input from all authors.

\bibliography{VIA.bib}

@article{Krizan_2026,
doi = {10.1088/2633-4356/ae7fe1},
url = {https://doi.org/10.1088/2633-4356/ae7fe1},
year = {2026},
month = {jul},
publisher = {IOP Publishing},
volume = {6},
number = {3},
pages = {036001},
author = {Križan, Christian and Toselli, Maurizio and Ahmad, Irshad and Khaksaran, Hadi and Rommel, Marcus and Trnjanin, Nermin and Biznárová, Janka and Dahiya, Mamta and Hogedal, Emil and Jakobsson, Halldór and Nylander, Andreas and Bylander, Jonas and Delsing, Per and Tancredi, Giovanna},
title = {Electrical post-fabrication tuning of aluminum Josephson junctions at room temperature},
journal = {Materials for Quantum Technology}
}

@article{chen2026unlocking,
  title={Unlocking a fast adiabatic CZ gate and exact residual $ ZZ $ cancellation between fixed-frequency transmons using a floating tunable coupler},
  author={Chen, Angela Q and Wu, Xian and Strong, Sarah and Poletto, Stefano},
  journal={arXiv preprint arXiv:2604.05048},
  year={2026}
}

@article{field2024modular,
  title={Modular superconducting-qubit architecture with a multichip tunable coupler},
  author={Field, Mark and Chen, Angela Q and Scharmann, Ben and Sete, Eyob A and Oruc, Feyza and Vu, Kim and Kosenko, Valentin and Mutus, Joshua Y and Poletto, Stefano and Bestwick, Andrew},
  journal={Physical Review Applied},
  volume={21},
  number={5},
  pages={054063},
  year={2024},
  publisher={APS}
}

@article{zeng2015direct,
  title={Direct observation of the thickness distribution of ultra thin AlO x barriers in Al/AlO x/Al Josephson junctions},
  author={Zeng, LJ and Nik, S and Greibe, Tine and Krantz, Philip and Wilson, CM and Delsing, Per and Olsson, Eva},
  journal={Journal of Physics D: Applied Physics},
  volume={48},
  number={39},
  pages={395308},
  year={2015},
  publisher={IOP Publishing}
}

@article{weeden2003probe,
  title={Probe card tutorial},
  author={Weeden, Otto},
  journal={Keithley Instruments, Inc},
  pages={1--40},
  year={2003}
}

@article{iaia2025non,
  title={Non-equilibrium Dynamics of Two-level Systems directly after Cryogenic Alternating Bias},
  author={Iaia, V and Joseph, ES and Im, S and Hagopian, N and O'Kelley, S and Kim, C and Materise, N and Patra, S and Lordi, V and Eriksson, MA and others},
  journal={arXiv preprint arXiv:2509.19223},
  year={2025}
}

@article{budoyo2026characterization,
  title={Characterization of Josephson Junction Aging and Annealing Under Different Environments},
  author={Budoyo, Rangga P and Kajen, Rasanayagam S and Cheah, Bing Wen and Nguyen, Long H and Dumke, Rainer},
  journal={arXiv preprint arXiv:2602.23888},
  year={2026}
}

@article{smirnov2025subangstrom,
  title={Subangstrom ion beam engineering of buried ultrathin oxides for scalable quantum computing},
  author={Smirnov, Nikita S and Krivko, Elizaveta A and Moskaleva, Daria A and Moskalev, Dmitry O and Solovieva, Anastasia A and Matanin, Aleksei R and Echeistov, Vladimir V and Ivanov, Anton I and Malevannaya, Elizaveta I and Polozov , Viktor I and others},
  journal={Science Advances},
  volume={11},
  number={19},
  pages={eads9744},
  year={2025},
  publisher={American Association for the Advancement of Science}
}

@article{xkqs-grnd,
  title = {Josephson Junction Tuning Described by Depinning Physics},
  author = {Kennedy, Oscar W. and Cole, Jared H. and Shelly, Connor D.},
  journal = {Phys. Rev. Lett.},
  volume = {135},
  issue = {19},
  pages = {196202},
  numpages = {7},
  year = {2025},
  month = {Nov},
  publisher = {American Physical Society},
  doi = {10.1103/xkqs-grnd},
  url = {https://link.aps.org/doi/10.1103/xkqs-grnd}
}

@article{mohseni2024build,
  title={How to build a quantum supercomputer: Scaling from hundreds to millions of qubits},
  author={Mohseni, Masoud and Scherer, Artur and Johnson, K Grace and Wertheim, Oded and Otten, Matthew and Aadit, Navid Anjum and Alexeev, Yuri and Bresniker, Kirk M and Camsari, Kerem Y and Chapman, Barbara and others},
  journal={arXiv preprint arXiv:2411.10406},
  year={2024}
}

@article{saadatmand2024superconducting,
  title={Superconducting qubits at the utility scale: The potential and limitations of modularity},
  author={Saadatmand, SN and Wilson, Tyler L and Hodson, Mark J and Field, Mark and Devitt, Simon J and Vijayan, Madhav Krishnan and Robertson, Alan and Le, Thinh P and Ruh, Jannis and Paler, Alexandru and others},
  journal={arXiv preprint arXiv:2406.06015},
  year={2024}
}

@inproceedings{wang2024precision,
  title={Precision frequency tuning of tunable transmon qubits using alternating-bias assisted annealing},
  author={Wang, Xiqiao and Howard, Joel and Sete, Eyob A and Stiehl, Greg and Kopas, Cameron and Poletto, Stefano and Wu, Xian and Field, Mark and Sharac, Nicholas and Eckberg, Christopher and others},
  booktitle={2024 IEEE International Conference on Quantum Computing and Engineering (QCE)},
  volume={1},
  pages={1315--1323},
  year={2024},
  organization={IEEE}
}

@article{Sete_para_2021,
  title = {Parametric-Resonance Entangling Gates with a Tunable Coupler},
  author = {Sete, Eyob A. and Didier, Nicolas and Chen, Angela Q. and Kulshreshtha, Shobhan and Manenti, Riccardo and Poletto, Stefano},
  journal = {Phys. Rev. Appl.},
  volume = {16},
  issue = {2},
  pages = {024050},
  numpages = {13},
  year = {2021},
  month = {Aug},
  publisher = {American Physical Society},
  doi = {10.1103/PhysRevApplied.16.024050},
  url = {https://link.aps.org/doi/10.1103/PhysRevApplied.16.024050}
}

@PREAMBLE{
 "\providecommand{\noopsort}[1]{}" 
 # "\providecommand{\singleletter}[1]{#1}%" 
}

@article{FritzStructureMechancialAlOxPhysRevMater2019,
  title = {Structural and nanochemical properties of AlOx Layers in Al/AlOx/Al-layer systems for Josephson junctions},
  volume = {3},
  ISSN = {2475-9953},
  url = {http://dx.doi.org/10.1103/PhysRevMaterials.3.114805},
  DOI = {10.1103/physrevmaterials.3.114805},
  number = {11},
  journal = {Physical Review Materials},
  publisher = {American Physical Society (APS)},
  author = {Fritz,  S. and Radtke,  L. and Schneider,  R. and Luysberg,  M. and Weides,  M. and Gerthsen,  D.},
  year = {2019},
  month = nov 
}

@article{WeigelHRTEMJofPhysCondMat2008,
  title = {High-resolution Al L2, 3-edge x-ray absorption near edge structure spectra of Al-containing crystals and glasses: coordination number and bonding information from edge components},
  volume = {20},
  ISSN = {1361-648X},
  url = {http://dx.doi.org/10.1088/0953-8984/20/13/135219},
  DOI = {10.1088/0953-8984/20/13/135219},
  number = {13},
  journal = {Journal of Physics: Condensed Matter},
  publisher = {IOP Publishing},
  author = {Weigel,  C and Calas,  G and Cormier,  L and Galoisy,  L and Henderson,  G S},
  year = {2008},
  month = mar,
  pages = {135219}
}

@article{LasiQBerkeley2022,
    author = {Kim, Hyunseong and Jünger, Christian and Morvan, Alexis and Barnard, Edward S. and Livingston, William P. and Altoé, M. Virginia P. and Kim, Yosep and Song, Chengyu and Chen, Larry and Kreikebaum, John Mark and Ogletree, D. Frank and Santiago, David I. and Siddiqi, Irfan},
    title = "{Effects of laser-annealing on fixed-frequency superconducting qubits}",
    journal = {Applied Physics Letters},
    volume = {121},
    number = {14},
    pages = {142601},
    year = {2022},
    month = {10},
    issn = {0003-6951},
    doi = {10.1063/5.0102092},
    url = {https://doi.org/10.1063/5.0102092},
    eprint = {https://pubs.aip.org/aip/apl/article-pdf/doi/10.1063/5.0102092/16484380/142601\_1\_online.pdf},
}

@Article{Hertzberg2021,
author={Hertzberg, Jared B.
and Zhang, Eric J.
and Rosenblatt, Sami
and Magesan, Easwar
and Smolin, John A.
and Yau, Jeng-Bang
and Adiga, Vivekananda P.
and Sandberg, Martin
and Brink, Markus
and Chow, Jerry M.
and Orcutt, Jason S.},
title={Laser-annealing Josephson junctions for yielding scaled-up superconducting quantum processors},
journal={npj Quantum Information},
year={2021},
month={Aug},
day={19},
volume={7},
number={1},
pages={129},
issn={2056-6387},
doi={10.1038/s41534-021-00464-5},
url={https://doi.org/10.1038/s41534-021-00464-5}
}

@article{LasiQScience2022,
author = {Eric J. Zhang  and Srikanth Srinivasan  and Neereja Sundaresan  and Daniela F. Bogorin  and Yves Martin  and Jared B. Hertzberg  and John Timmerwilke  and Emily J. Pritchett  and Jeng-Bang Yau  and Cindy Wang  and William Landers  and Eric P. Lewandowski  and Adinath Narasgond  and Sami Rosenblatt  and George A. Keefe  and Isaac Lauer  and Mary Beth Rothwell  and Douglas T. McClure  and Oliver E. Dial  and Jason S. Orcutt  and Markus Brink  and Jerry M. Chow },
title = {High-performance superconducting quantum processors via laser annealing of transmon qubits},
journal = {Science Advances},
volume = {8},
number = {19},
pages = {eabi6690},
year = {2022},
doi = {10.1126/sciadv.abi6690},
URL = {https://www.science.org/doi/abs/10.1126/sciadv.abi6690},
eprint = {https://www.science.org/doi/pdf/10.1126/sciadv.abi6690}}

@article{Kreikebaum_2020,
    doi = {10.1088/1361-6668/ab8617},
    url = {https://dx.doi.org/10.1088/1361-6668/ab8617},
    year = {2020},
    month = {apr},
    publisher = {IOP Publishing},
    volume = {33},
    number = {6},
    pages = {06LT02},
    author = {J M Kreikebaum and K P O’Brien and A Morvan and I Siddiqi},
    title = {Improving wafer-scale Josephson junction resistance variation in superconducting quantum coherent circuits},
    journal = {Superconductor Science and Technology}
}

@misc{korshakov2024aluminum,
      title={Aluminum Josephson junction microstructure and electrical properties modification with thermal annealing}, 
      author={N. D. Korshakov and D. O. Moskalev and A. A. Soloviova and D. A. Moskaleva and E. S. Lotkov and A. R. Ibragimov and M. V. Androschuk and I. A. Ryzhikov and Y. V. Panfilov and I. A. Rodionov},
      year={2024},
      eprint={2403.02179},
      archivePrefix={arXiv},
      primaryClass={quant-ph}
}

@misc{balaji2024electronbeam,
      title={Electron-beam annealing of Josephson junctions for frequency tuning of quantum processors}, 
      author={Yashwanth Balaji and Narendra Acharya and Robert Armstrong and Kevin G. Crawford and Sergey Danilin and Thomas Dixon and Oscar W. Kennedy and Renuka Devi Pothuraju and Kowsar Shahbazi and Connor D. Shelly},
      year={2024},
      eprint={2402.17395},
      archivePrefix={arXiv},
      primaryClass={quant-ph}
}

@misc{pappas2024alternating,
      title={Alternating Bias Assisted Annealing of Amorphous Oxide Tunnel Junctions}, 
      author={David P. Pappas and Mark Field and Cameron Kopas and Joel A. Howard and Xiqiao Wang and Ella Lachman and Lin Zhou and Jinsu Oh and Kameshwar Yadavalli and Eyob A. Sete and Andrew Bestwick and Matthew J. Kramer and Joshua Y. Mutus},
      year={2024},
      eprint={2401.07415},
      archivePrefix={arXiv},
      primaryClass={physics.app-ph}
}

@article{PhysRevLett.10.486,
  title = {Tunneling Between Superconductors},
  author = {Ambegaokar, Vinay and Baratoff, Alexis},
  journal = {Phys. Rev. Lett.},
  volume = {10},
  issue = {11},
  pages = {486--489},
  numpages = {0},
  year = {1963},
  month = {Jun},
  publisher = {American Physical Society},
  doi = {10.1103/PhysRevLett.10.486},
  url = {https://link.aps.org/doi/10.1103/PhysRevLett.10.486}
}

@article{gold2021entanglement,
  title={Entanglement across separate silicon dies in a modular superconducting qubit device},
  author={Gold, Alysson and Paquette, JP and Stockklauser, Anna and Reagor, Matthew J and Alam, M Sohaib and Bestwick, Andrew and Didier, Nicolas and Nersisyan, Ani and Oruc, Feyza and Razavi, Armin and others},
  journal={npj Quantum Information},
  volume={7},
  number={1},
  pages={142},
  year={2021},
  publisher={Nature Publishing Group UK London}
}

@article{PhysRevA.80.052312,
  title = {High-threshold universal quantum computation on the surface code},
  author = {Fowler, Austin G. and Stephens, Ashley M. and Groszkowski, Peter},
  journal = {Phys. Rev. A},
  volume = {80},
  issue = {5},
  pages = {052312},
  numpages = {14},
  year = {2009},
  month = {Nov},
  publisher = {American Physical Society},
  doi = {10.1103/PhysRevA.80.052312},
  url = {https://link.aps.org/doi/10.1103/PhysRevA.80.052312}
}

@article{granata2007localized,
  title={Localized laser trimming of critical current in niobium based Josephson devices},
  author={Granata, C and Vettoliere, A and Petti, L and Rippa, M and Ruggiero, B and Mormile, P and Russo, M},
  journal={Applied physics letters},
  volume={90},
  number={23},
  year={2007},
  publisher={AIP Publishing}
}

@article{pishchimova2023improving,
  title={Improving Josephson junction reproducibility for superconducting quantum circuits: junction area fluctuation},
  author={Pishchimova, Anastasiya A and Smirnov, Nikita S and Ezenkova, Daria A and Krivko, Elizaveta A and Zikiy, Evgeniy V and Moskalev, Dmitry O and Ivanov, Anton I and Korshakov, Nikita D and Rodionov, Ilya A},
  journal={Scientific Reports},
  volume={13},
  number={1},
  pages={6772},
  year={2023},
  publisher={Nature Publishing Group UK London}
}

@article{osman2021simplified,
  title={Simplified Josephson-junction fabrication process for reproducibly high-performance superconducting qubits},
  author={Osman, A and Simon, J and Bengtsson, A and Kosen, S and Krantz, P and P Lozano, D and Scigliuzzo, M and Delsing, P and Bylander, Jonas and Fadavi Roudsari, A},
  journal={Applied Physics Letters},
  volume={118},
  number={6},
  year={2021},
  publisher={AIP Publishing}
}

@article{migacz2003thermal,
  title={Thermal annealing of Nb/Al-AlO/sub x//Nb Josephson junctions},
  author={Migacz, Justin V and Huber, Martin E},
  journal={IEEE transactions on applied superconductivity},
  volume={13},
  number={2},
  pages={123--126},
  year={2003},
  publisher={IEEE}
}

@article{koppinen2007complete,
  title={Complete stabilization and improvement of the characteristics of tunnel junctions by thermal annealing},
  author={Koppinen, PJ and V{\"a}ist{\"o}, LM and Maasilta, IJ},
  journal={Applied physics letters},
  volume={90},
  number={5},
  year={2007},
  publisher={AIP Publishing}
}
\clearpage
\appendix

\renewcommand{\thefigure}{S\arabic{figure}}
\renewcommand{\thetable}{S\arabic{table}}
\renewcommand{\theequation}{S\arabic{equation}}
\setcounter{figure}{0}
\setcounter{table}{0}
\setcounter{equation}{0}

\section*{Supplementary Material}

\section{Frequency Prediction Method}

The frequency prediction framework employed in this study adapts prior methodologies \cite{LasiQScience2022,Hertzberg2021,wang2024precision} that parameterize the qubit frequency as \(f \approx \beta E_J^{\alpha}\), where \(\beta\) and \(\alpha\) are derived empirically. Specifically,

\begin{enumerate}
    \item For each qubit in the QPU design, we extract the simulated charging energy, \(E_{C,\mathrm{sim}}\), via electromagnetic simulation. An internal numerical solver then uses \(E_{C,\mathrm{sim}}\) and the design Hamiltonian to compute the dressed qubit frequency, \(f_{\mathrm{sim}}\), and the corresponding Josephson energy, \(E_{J,\mathrm{sim}}\).

    \item Using the empirical Ambegaokar--Baratoff relation \cite{PhysRevLett.10.486}, we estimate the Josephson energy from the room-temperature normal-state junction resistance, \(R_N\), and and the corresponding coefficient, \(\alpha_{\mathrm{AB}}\), as
    \begin{equation}
        E_{J,\mathrm{est}}
        =
        \frac{\alpha_{\mathrm{AB}}}{R_N}.
    \end{equation}

    \item Using the simulated charging energy from the QPU layout and the transmon approximation assuming small \(E_{C}\), the qubit frequency is estimated as
    \begin{equation}
        f_{\mathrm{pred}}
        \approx
        f_{\mathrm{sim}}
        \sqrt{
            \frac{E_{J,\mathrm{est}}}
                 {E_{J,\mathrm{sim}}}
        }.
    \end{equation}
\end{enumerate}

In our model, the transmon approximation sets \(\alpha=1/2\), while the prefactor \(\beta=f_{\mathrm{sim}}/E_{J,\mathrm{sim}}^{1/2}\), is derived from simulations. Systematic offset errors in the predicted frequency—such as the $132~\mathrm{MHz}$ mean deviation of measured $f_{01}^{\max}$ values from their design targets (Table~\ref{tab:long})—likely stem from finite accuracies in the simulated shunt capacitance and wafer-to-wafer variations in the empirical Ambegaokar--Baratoff coefficient, $\alpha_{\mathrm{AB}}$. Changes in junction resistance during post-ABAA processing and thermal cycling may also contribute to this offset. To enhance prediction accuracy for future large-scale, high-performance QPUs, we are currently refining our electromagnetic simulations and investigating methods to better control \(\alpha_{\mathrm{AB}}\) alongside the impacts of post-ABAA processing.

\begin{figure*}
\includegraphics[width=1.0\textwidth]{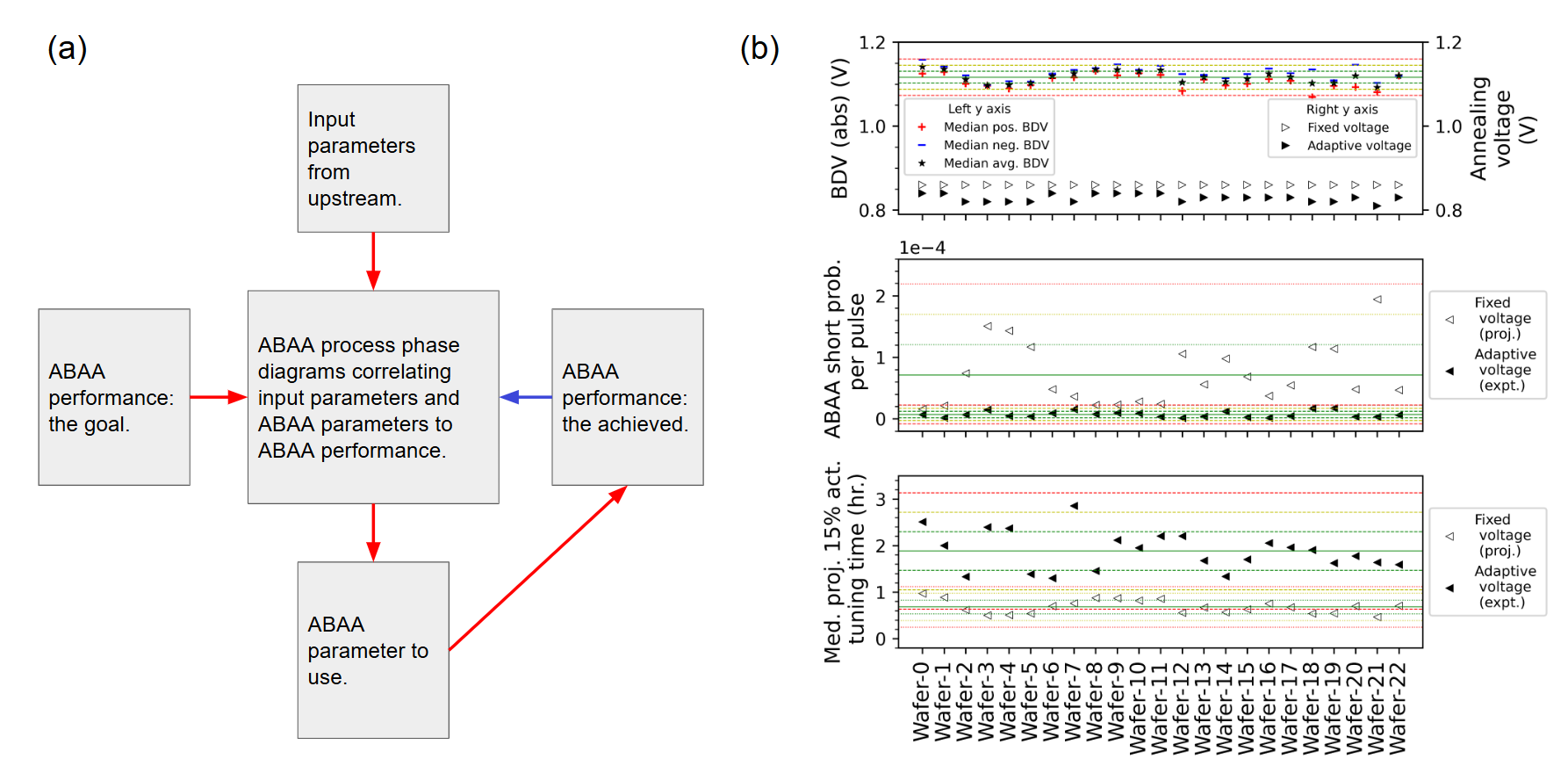}
\caption{ABAA process control. (a) A combined feedforward (red arrows) and feedback (blue arrow) strategy for ABAA process control. (b) An example of ABAA process control to achieve a short probability of $\leq 2\times 10^{-5}$/pulse utilizing the phase diagram in Fig. 3 of the main text. Upper panel: Wafer-median breakdown voltages are plotted on the left y-axis: positive BDV (red crosses), negative BDV (blue minus signs), and their average (black stars). The horizontal solid line represents the mean, while the dashed lines indicate the 1-, 2-, and 3-$\sigma$ variation intervals of the black star data points. Experimentally applied ABAA annealing voltage (solid triangle, adaptive to BDV variations) and a fixed annealing voltage of 0.86V (empty triangle for comparison) are plotted on the right y-axis. Middle panel: Solid triangles show the experimentally measured ABAA short probability per pulse. Empty triangles represent the short probability extrapolated from the 0.86 V calibration fit (teal line, Fig. 3(a)). Bottom panel: Solid (empty) triangles indicate the experimental (extrapolated) tuning time corresponding to the adaptively applied (fixed) annealing voltage. In the middle and bottom panels, the horizontal solid line represents the mean, while the dashed (dotted) lines indicate the 1-, 2-, and 3-$\sigma$ variation intervals of the solid (empty) data points.
}
\label{Fig:FigureS1}
\end{figure*}

\section{ABAA process control}

The primary objective of ABAA process control is to meet ABAA performance targets—such as throughput and yield—dictated by foundry production delivery metrics. As illustrated in Figure S1(a), the methodology introduced here utilizes a combined feedforward and feedback control strategy, centered around a multi-dimensional process control phase diagram, such as the one showin in Fig. 3. The phase diagram correlates upstream input parameters (physical properties, e.g., JJ BDV and RA-product, of the target devices to be tuned by ABAA) and ABAA process parameters (control knobs of the ABAA process, e.g., annealing voltage, temperature) with final performance outcomes (e.g., ABAA throughput, yield, targeting). Establishing optimal process performance goals typically requires balancing competing factors, such as tuning speed and short probability, or prioritizing one over the other based on specific production requirements. 

The feedforward control loop measures input parameters and proactively adjusts ABAA process variables to compensate for incoming offsets and variations, ensuring the output meets performance goals while mitigating disturbances from upstream. Conversely, the feedback control loop evaluates actual ABAA performance against these goals. This acts as a diagnostic tool to fine-tune the control phase diagram and optimize both ABAA and upstream processes for future stability.  

Supplementary Fig. S1(b) illustrates an example of the ABAA process control methodology that implements the phase diagram from Fig. 3 of the main text. It demonstrates effective process control by adaptively selecting annealing voltages based on the measured BDV of incoming JJ devices, achieving the ABAA yield performance target of $\leq 2\times 10^{-5}$/pulse while maintaining predictable process throughput. In the upper panel of Fig. S1, the wafer-median BDV and annealing voltage are plotted on the left and right axes, respectively. The average of the median positive and negative BDVs is $1116 \pm 14$ mV across wafers. Guided by the short ratio vs. BDV phase diagram in Fig. 3(a), the applied annealing voltage (solid triangles) is selected with a safety margin and adaptively adjusted in 10 mV steps based on median BDV variations. For comparison, a fixed annealing voltage of 0.86 V—chosen based on an initial guess without relying on the phase diagram—is plotted to illustrate the effect of feedforward control on the annealing voltage.

The middle and bottom panels of Fig. S1(b) illustrate the ABAA short probability and tuning speed performance. Solid triangles denote the experimentally measured performance under the applied adaptive voltage, while empty triangles represent the projected performance for a fixed 0.86 V bias (extrapolated from the calibration fit in Fig. 3(a)). Although this 0.86 V projection meets the performance goal for wafer-0, it quickly deviates and fails the yield target as the BDV drops, exhibiting variations that strongly correlate with wafer-to-wafer BDV fluctuations.

In contrast, the applied voltage is intentionally lowered to provide a buffer against process variations and calibration errors, and it is proactively adjusted in response to BDV fluctuations. This approach yields a tightly distributed short probability of $(7 \pm 5) \times 10^{-6}$/pulse, where the upper edge of the 3-$\sigma$ limit ($2.2 \times 10^{-5}$/pulse) closely aligns with the target upper bound. Because the annealing voltage dictates a trade-off between short probability and tuning time, the longer tuning time required by the adaptive voltage represents a necessary compromise to prioritize yield (i.e., short probability). Ultimately, both the measured ABAA performance data and the incoming wafer BDVs serve as direct feedback to calibrate the ABAA phase diagram and refine upstream JJ processes, thereby enhancing overall control stability and predictability.

\end{document}